\documentclass[prl,twocolumn,amsmath,amssymb,showpacs]{revtex4}

\usepackage{graphicx}
\usepackage{dcolumn}
\usepackage{bm}
\usepackage{mathrsfs}
\usepackage{appendix}
\usepackage{bbm}
\usepackage{color}

\def \vS {{\bf S}}
\def \vR {{\bf R}}
\def \ve {{\bf e}}
\def \vQ {{\bf Q}}
\def \vh {{\bf h}}
\def \sgn {{\rm sgn}}
\def \pin {{\bf P}^{\rm ind} }
\def \xp {{\bf x}^{\prime }}
\def \yp {{\bf y}^{\prime }}
\def \zp {{\bf z}^{\prime }}

\begin{document}

\title{Identifying the Magnetoelectric Modes of Multiferroic BiFeO$_3$}

\author{Randy S. Fishman$^1$, Nobuo Furukawa$^2$, Jason T. Haraldsen$^{3,4}$, Masaaki Matsuda$^5$, and Shin Miyahara$^6$}

\affiliation{$^1$Materials Science and Technology Division, Oak Ridge National Laboratory, Oak Ridge, Tennessee 37831, USA}
\affiliation{$^2$Department of Physics and Mathematics, Aoyama Gakuin University, Sagamihara, Kanagawa 229-8558, Japan}
\affiliation{$^3$Theoretical Division, Los Alamos National Laboratory, Los Alamos, New Mexico 87545, USA}
\affiliation{$^4$Center for Integrated Nanotechnologies, Los Alamos National Laboratory, Los Alamos, New Mexico 87545, USA}
\affiliation{$^5$Quantum Condensed Matter Division, Oak Ridge National Laboratory, Oak Ridge, Tennessee 37831, USA}
\affiliation{$^6$Asia Pacific Center for Theoretical Physics, Pohang University of Science and Technology, Pohang, Gyeongbuk, 790-784, Korea}
\date{\today}

\begin{abstract}

We have identified three of the four magnetoelectric modes of multiferroic BiFeO$_3$ measured using THz spectroscopy.  Excellent agreement with the 
observed peaks is obtained by including the effects of easy-axis anisotropy along the direction of the electric polarization.  By distorting the cycloidal spin state, 
anisotropy splits the $\Psi_{\pm 1}$ mode into peaks at 20 and 21.5 cm$^{-1}$ and activates the lower $\Phi_{\pm 2}$ mode at 27 cm$^{-1}$ ($T=200$ K).  
An electromagnon is identified with the upper $\Psi_{\pm 1}$ mode at 21.5 cm$^{-1}$.  Our results also explain recent Raman and
inelastic neutron-scattering measurements.

\end{abstract}


\maketitle

Multiferroic materials hold tremendous technological promise due to the coupling between the electric polarization and cycloidal magnetic order.  
Two classes of multiferroic materials have been identified.  In ``proper" multiferroics, the cycloid develops at a lower temperature than the ferroelectric polarization;  
in ``improper" multiferroics, the electric polarization is directly coupled to the cycloid \cite{khomskii06, cheong07} so they develop at the same temperature.  
Although the multiferroic coupling is typically stronger in ``improper" multiferroics, those materials also have rather low transition temperatures.  Due to the high magnetic transition temperatures of ``proper" multiferroics like BiFeO$_3$ with $T_{\rm N} \approx 640$ K
\cite{sosnowska82, lebeugle08, slee08}, those materials continue to attract a great deal of interest.  With the availability of single crystals, the physical understanding of BiFeO$_3$
has advanced rapidly over the past few years but the description of the distorted spin cycloid and microscopic interactions in BiFeO$_3$ remains incomplete.  

Recent inelastic neutron-scattering measurements on single-crystal samples of BiFeO$_3$ \cite{jeong12, matsuda12} were used to estimate the easy-axis anisotropy $K$ along the polarization direction as well as the Dzyaloshinskii-Moriya (DM) coupling 
$D$ that induces the long-period cycloidal order with wavevector $\vQ = (2\pi /a)(0.5+\delta ,0.5, 0.5-\delta )$ and $\delta \approx 0.0045$ \cite{ramazanoglu11, herrero10, sosnowska11}.   However, more precise estimates for those
coupling constants may be obtained based on the magnetoelectric peaks recently measured using THz spectroscopy \cite{talbayev11}.  In this work, we evaluate the magnetoelectric peaks from a model that includes 
easy-axis anisotropy and DM interactions.  In addition to providing excellent agreement for three of the four experimental peaks, we also provide several new predictions for the selection rules governing those three peaks and one higher-energy peak.

\begin{figure}
\includegraphics[width=7.0cm]{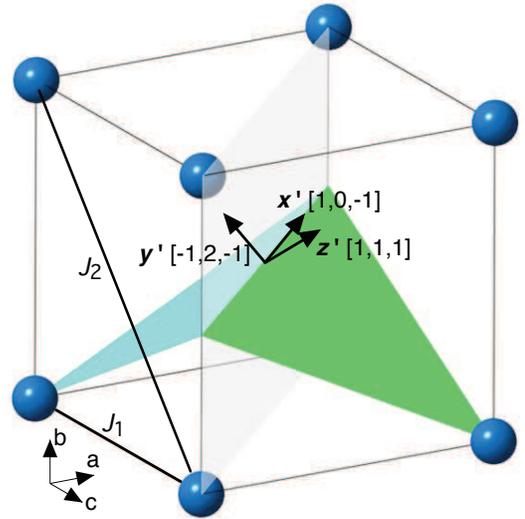}
\caption{(Color online)  The pseudo-cubic unit cell for BiFeO$_3$ showing $\xp $, $\yp $, and $\zp $ directions.
The distorted spin spiral propagates along the $\xp $ direction with spins in the $(-1,2,-1)$ plane.  AF interactions $J_1$ and $J_2$ are also indicated. 
}
\label{structure}
\end{figure}

Due to the displacement of the Bi$^{3+}$ ions, ferroelectricity appears in rhombohedral BiFeO$_3$ below $T_c \approx 1100$ K \cite{teague70}.
For the pseudo-cubic unit cell in Fig.1 with lattice constant $a\approx 3.96 \AA $, 
cycloidal order develops below $T_{\rm N}$ with propagation vector $(2\pi /a)(\delta  ,0,-\delta )$ along 
$\xp $ for each $(1,1,1)$ plane  \cite{sosnowska82,lebeugle08,slee08}.  Neighboring $(1,1,1)$ planes 
are coupled by the antiferromagnetic (AF) interaction $J_1$ between the $S=5/2$ Fe$^{3+}$ spins.   
The DM interaction ${\bf D}$ along $\yp $ or $[-1,2,-1]$ produces a cycloid with spins in the $(-1,2,-1)$ plane.  Easy-axis anisotropy $K$ 
along $[1,1,1]$, parallel to the polarization ${\bf P}$, distorts the cycloid by producing odd harmonics of the fundamental ordering wavevector \cite{kadomtseva04}.

Although $T_{\rm N}$ is much lower than $T_c$, the magnetic domain distribution of BiFeO$_3$ can be effectively manipulated 
by an electric field \cite{lebeugle08,slee08,slee08_2}.  In a magnetic field above 20 T, the transformation of the cycloid to an almost commensurate structure with 
weak ferromagnetic moment \cite{ohoyama11} is accompanied by a sharp drop of about 40 nC/cm$^2$ in the electric polarization \cite{kadomtseva04,park11,tokunaga10}.  
Therefore, the additional polarization $\pin $ observed below $T_{\rm N}$ is induced by the cycloidal spin state.  

This coupling between the cycloid and electric polarization is produced by the inverse DM mechanism \cite{katsura05, mostovoy06, sergienko06}
with induced polarization $\pin \propto \ve_{ij}\times (\vS_i\times \vS_j)$ where $\ve_{ij}=\vR_j-\vR_i$ and $\vS_i$ are the Fe$^{3+}$ spins.
Within each $(1,1,1)$ plane, $\ve_{ij}=\sqrt{2}a\xp $ connects spins at $\vR_i$ and 
$\vR_j =\vR_i+\sqrt{2}a\xp $.  So $\vS_i \times \vS_j$ 
points  along $\yp $ and $\pin $ points along $\zp $.

To evaluate the spin-wave (SW) excitations of BiFeO$_3$, 
we use the Hamiltonian \cite{talbayev11,sosnowska95, jeong12}:
\begin{eqnarray}
&&H = J_1\sum_{\langle i,j\rangle }\mbox{\boldmath $S$}_i\cdot\mbox{\boldmath $S$}_j +J_2\sum_{\langle i,j \rangle'} \mbox{\boldmath $S$}_i\cdot\mbox{\boldmath $S$}_j
-K\sum_i (\vS_i\cdot {\bf z'})^2
\nonumber \\
&&-{D\sum }_{\vR_j=\vR_i +\sqrt{2}a\xp}\,\yp \cdot (\mbox{\boldmath $S$}_{i}\times\mbox{\boldmath $S$}_j). 
\label{Ham}
\end{eqnarray}
In the first and second exchange terms, $\langle i,j\rangle $ denotes a sum over nearest neighbors and $\langle i,j\rangle'$ a sum over next-nearest neighbors.
The third term originates from the easy-axis anisotropy along $\zp $ and the fourth term from the DM interaction with ${\bf D}$ along $\yp $.  
For a fixed cycloidal period, $D$ is a 
smoothly increasing function of  $K$, as shown in Fig.5 of Ref.\cite{matsuda12}.  

Since $\delta = 0.0045$ is close to 1/222, a unit cell containing 222 sites within each $(1,1,1)$ plane was used to characterize the distorted cycloid, 
which was expanded in odd harmonics of the fundamental wavevector $\vQ =(2\pi /a)(0.5+\delta, 0.5, 0.5-\delta )$ \cite{fishman10}:
\begin{equation}
\label{sp1}
S_{z^{\prime }}(\vR )=S \sum_{m=0}^{\infty } C_{2m+1} \cos \bigl((2m+1)\vQ \cdot \vR \bigr),
\end{equation}
\begin{equation}
\label{sp2}
S_{x^{\prime }}(\vR ) = \sqrt{S^2-S_{z'}(\vR )^2}\, \sgn \bigl(\sin (\vQ \cdot \vR ) \bigr).
\end{equation}
The harmonic coefficients obey the sum $\sum_{m=0}^{\infty }C_{2m+1} =1$.

\begin{figure}
\includegraphics[width=8.8cm]{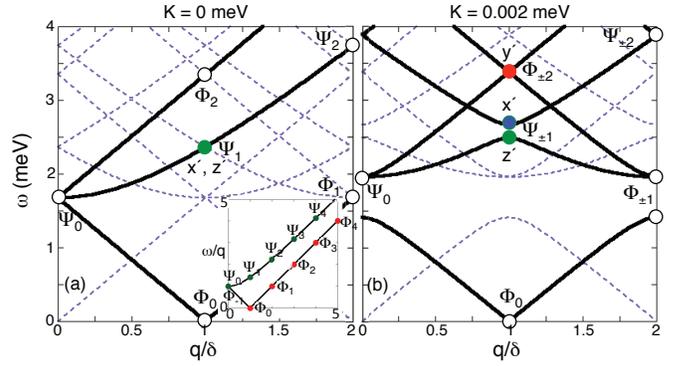}
\caption{(Color online)  The SW modes of BiFeO$_3$ versus $q/\delta $ for wavevector $(2\pi /a)(0.5+q,0.5,0.5-q)$.  Dashed lines show all possible excitations and 
the solid lines show only those modes with non-zero intensity.  (a) For $K=0$, the two strongest SW branches are plotted versus $q/\delta $ in the inset with points $\Phi_n$ and $\Psi_n$
at multiples of $\delta $ for $c=1$.  The solid (green) point at $\Psi_1$ indicates a MR and EM mode.  
(b) For $K=0.002$ meV, the $\Psi_{\pm 1}$ mode (green and blue) splits and the lower $\Phi_{\pm 2}$ mode (red) is activated.   The $x^{\prime }$ (blue), $y^{\prime }$ (red), and 
$z^{\prime }$ (green) components of the MR modes are indicated.  The EM mode corresponds to the upper (blue) $\Psi_{\pm 1}$ mode}
\label{SWs}
\end{figure}

Excitation frequencies and intensities were evaluated by performing a $1/S$ expansion in the rotated frame of reference for each spin \cite{haraldsen09}
in the 444-site unit cell with two AF-coupled layers.  
Due to zone folding, there are 222 positive eigenfrequencies for each wavevector.  However, only a handful of those frequencies have any intensity.  

To evaluate the SW frequencies, we use AF interactions $J_1=4.5$ meV and $J_2=0.2$ meV, which describe the inelastic neutron-scattering 
measurements \cite{jeong12, matsuda12} at 200 K \cite{temp}.  Below 4 meV and for $0 \le q < 2\delta $, all possible SW frequencies are denoted by the dashed curves
in Fig.2 for $K=0$ and 0.002 meV.  Branches with nonzero intensity are indicated by the dark solid curves.   

When $K=0$, the two strongest SW branches are plotted in the inset to Fig.2(a) with points at multiples of the wavevector $q=\delta $ labeled as $\Phi_n$ or $\Psi_n$.
For small frequencies and $q=n\delta $, $\omega (\Phi_{n-1}) = \vert n-1\vert \, c\delta  $ and $\omega (\Psi_n) = \sqrt{ 1 + n^2}\, c\delta $,
where $c$ is the SW velocity of the linear branches \cite{sousa08}.
In the reduced-zone scheme, odd-$n$ $\Psi_n$ and even-$n$ $\Phi_n$ modes lie at the zone center $q=\delta $ while
even-$n$ $\Psi_n$ and odd-$n$ $\Phi_n$ modes lie at the zone edge $q=0 $.

When $K > 0$, the SW spectrum for small frequencies changes dramatically.  Higher harmonics of the cycloid split every set of crossing 
$\Phi_{\pm n}$ and $\Psi_{\pm n}$ modes.  The largest splitting occurs at $q=0$ between
the $\Phi_{\pm 1}$ modes.  A smaller splitting occurs at $q=\delta $ between the $\Psi_{\pm 1}$ modes.  While the $\Phi_0$ mode is shifted slightly above zero frequency, 
the $\Psi_0$ mode is moved up to just below the top $\Phi_{\pm 1}$ mode.  Figure 3 plots the evolution of those points with anisotropy.  
Although too small to appear in this plot, even the $\Phi_{\pm 2}$ modes are split by anisotropy.

\begin{figure}
\includegraphics[width=8.0cm]{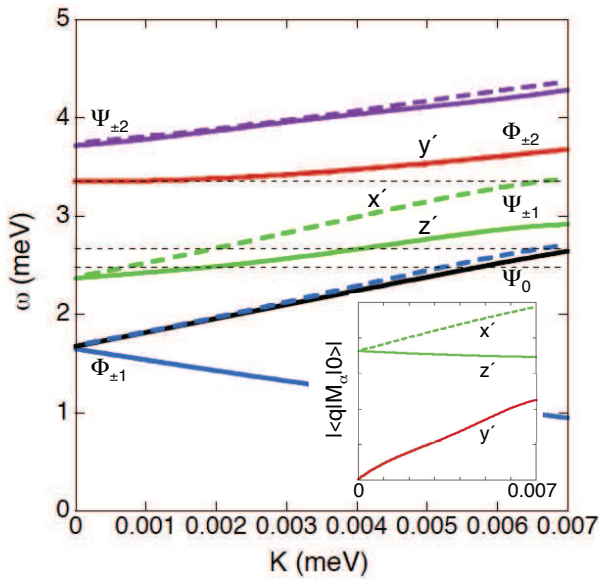}
\caption{(Color online) The evolution of the SW frequencies with easy-axis anisotropy.  Solid and dashed pairs of curves indicate split $\pm n$ modes.  
Inset are the matrix elements for the MR modes (arbitrary units) versus $K$.  The horizontal dashed lines indicate the 
experimental results of Ref.\cite{talbayev11} at 200 K.}
\end{figure}

Spectroscopic intensities are given by the matrix elements of $M_{\alpha }=g\mu_B \sum_i S_{i \alpha }$ for magnetic resonance (MR) and by the matrix elements of 
the induced polarization
\begin{equation}
P^{\rm ind}_{\alpha } \propto \sum_{\vR_j=\vR_i +\sqrt{2}a\xp } \Bigl\{ \xp \times (\vS_i \times \vS_j)\Bigr\}_{\alpha }
\end{equation}
for the electromagnon (EM) \cite{miyahara12}.  Each matrix element is evaluated between the ground state $\vert 0\rangle $ and an excited state $\vert q\rangle $ containing a single magnon with wavevector
at the cycloidal zone center $q=\delta $.

When $K=0$, there is a single MR peak at the $\Psi_1$ point in Fig.2(a).  For this mode,
$\vert \langle q \vert M_{x^{\prime }}\vert 0\rangle \vert = \vert \langle q \vert M_{z^{\prime }}\vert 0\rangle \vert > 0 $ and $\vert \langle q \vert M_{y'}\vert 0\rangle \vert = 0$.  
An EM peak with component $y^{\prime }$ coincides with the MR peak.
Only the $y'$ component of $\langle q \vert P^{\rm ind}_{\alpha }\vert 0\rangle $ is nonzero.  No other frequencies at $q=\delta $ are magnetoelectrically active for $K=0$.

When $K > 0$, both $\Psi_{\pm 1}$ modes are active with nonzero $x^{\prime }$ and $z^{\prime }$ MR matrix elements for the upper and lower modes, respectively.  The MR matrix elements 
$\vert \langle q \vert M_{\alpha }\vert 0\rangle \vert $ of those modes are plotted versus anisotropy in the inset to Fig.3.  
While the matrix element of the lower $\Psi_{\pm 1}$ mode decreases with anisotropy, that of the 
upper $\Psi_{\pm 1}$ mode increases.  Because the third harmonic of the $\Psi_{\pm 1}$ modes couples to the first harmonic of the $\Phi_{\pm 2}$ modes, 
the lower $\Phi_{\pm 2}$ mode is activated by anisotropy \cite{sousa08} with a nonzero $y'$ MR matrix element plotted in the inset to Fig.3.  

A large EM peak with matrix element $\langle q \vert P^{\rm ind}_{y^{\prime }}\vert 0\rangle $ coincides with the upper $\Psi_{\pm 1}$ mode.  Another EM peak with
matrix element about 10 times smaller was found at the upper $\Psi_{\pm 3}$ mode with frequency 5.39 meV (43.4 cm$^{-1}$).  
This peak also has a significant MR matrix element $\vert \langle q \vert M_{x^{\prime }}\vert 0\rangle \vert $.  Hence, the active $\Psi_{\pm (2m+1)}$ peaks 
correspond to out-of-plane modes of the cycloid (excited by magnetic fields
along $\xp $ or $\zp $) while the active $\Phi_{\pm 2m}$ peaks correspond to in-plane modes of the cycloid (excited by a magnetic field along $\yp $).

Below 30 cm$^{-1}$, THz spectroscopy \cite{talbayev11} observed four infrared modes with wavevectors 
(measured at or extrapolated to 200 K \cite{temp}) of 17.5, 20, 21.5, and 27 cm$^{-1}$.
The upper three mode frequencies are denoted by the dashed lines in Fig.3.  These modes can be quite accurately described by $K = 0.002$ meV,
which produces MR peaks at 2.49, 2.67, and 3.38 meV, remarkably close to the observed peaks at 2.48, 2.67, and 3.35 meV.  Talbayev {\em et al.} \cite{talbayev11}
conjectured that the 20 and 21.5 cm$^{-1}$ lines were produced by the splitting of the $\Psi_{\pm 1}$ modes due to a modulated DM interaction along $\zp $.
We conclude that the splitting of the $\Psi_{\pm 1}$ modes is caused by easy-axis anisotropy along ${\bf z'}$ with quantitatively accurate values.  
Perhaps due to the small matrix element plotted in the inset to Fig.3, the lower $\Phi_{\pm 2}$ peak at 27 cm$^{-1}$ was not detected \cite{talbayev11} above about 150 K.  

The selection rules governing the MR modes also agree with the results of Ref.\cite{talbayev11}.  Field directions $\vh_1  \| [1,-1,0]$ and
$\vh_2  \| [1,1,0]$ can be written as $\vh_1 = \xp /2-\sqrt{3}\yp /2$ and $\vh_2 = \xp /2+\sqrt{3}\yp /6 +\sqrt{2/3}\zp $.
Consequently, the lower $\Phi_{\pm 2}$ mode with MR component $y'$ and the upper $\Psi_{\pm 1}$ mode with MR component $x'$ are 
excited both by fields $\vh_1$ and $\vh_2$, but the lower $\Psi_{\pm 1}$ mode with MR component $z'$ is only excited by field 
$\vh_2$ \cite{magnetic-domain}.  The predicted upper $\Psi_{\pm 3}$ mode with MR component $x'$ should be excited by both fields $\vh_1$ and $\vh_2$.

Since it is associated with a large EM peak, the upper $\Psi_{\pm 1}$ mode can also be excited by an electric field along $\yp $.  The
observation of non-reciprocal directional dichroism (NDD) under an external magnetic field ${\bf B}^{\rm ex}$ along $\zp$ \cite{miyahara12}
would confirm this prediction.  NDD requires linearly-polarized electromagnetic waves propagating along 
$\xp $ with electric and magnetic components ${\bf E}^\omega \| \yp $ and ${\bf H}^\omega \| \zp $.

By contrast, the observed low-energy mode at wavevector 17.5 cm$^{-1}$ (2.17 meV) cannot be explained by our model 
because there are no zone-center excitations 
below the lower $\Psi_{\pm 1}$ mode at 2.5 meV.  Nevertheless, the 2.17 meV peak lies tantalizingly close to both the upper
$\Phi_{\pm 1}$ mode and the $\Psi_0$ mode at the zone boundary in Fig.2(b).  It is possible that the 2.17 meV peak is 
associated with a more complex magnetic structure induced by some interaction or anisotropy not included in our model.
Indeed, additional long-range order with wavevector $(2\pi /a)(\delta, 0, -\delta)$ or $(2\pi /a)(0.5,0.5,0.5)$ would
hybridize the $\Phi_{\pm 1}$ and $\Psi_0$ modes at the zone edge with the $\Psi_{\pm 1}$ modes at the zone center.
In Raman spectroscopy, the anomalies of the 2.17 meV peak at 140 K and 200 K were identified with spin reorientations \cite{cazayous08, singh08} of the cycloid.  
However, any spin reorientation must respect the selection rules for the zone-center MR modes.

All of the active zone-center modes have been observed with Raman spectroscopy \cite{cazayous08, singh08, rovillain09}.  
The Raman peak \cite{rovillain09} at  23 cm$^{-1}$ ($T = 10$ K) 
detected with parallel polarizations (exciting out-of-plane modes) can be identified with the 
$\Psi_{\pm 1}$  modes and the 25.5 cm$^{-1}$ peak detected with crossed polarizations (exciting in-plane modes)
can be identified with the lower $\Phi_{\pm 2}$ mode.  Another Raman peak \cite{singh08} at 43.4 cm$^{-1}$ ($T=80$ K) lies 
quite near the estimated wavevector of the upper $\Psi_{\pm 3}$ mode.

Because it is governed by different selection rules than THz spectroscopy, Raman spectroscopy has also observed several 
zone-edge cycloidal modes.  That includes strong peaks at 34.5 cm$^{-1}$ (4.28 meV, $T=80$ K) \cite{singh08} 
and 32 cm$^{-1}$ (3.96 meV, $T = 10$ K) \cite{rovillain09},
close to the predicted energy of the zone-edge $\Psi_{\pm 2}$ modes.  As conjectured above, additional long-range order may produce such 
zone-edge cycloidal peaks in the Raman spectrum.  

\begin{figure}
\includegraphics[width=8.5cm]{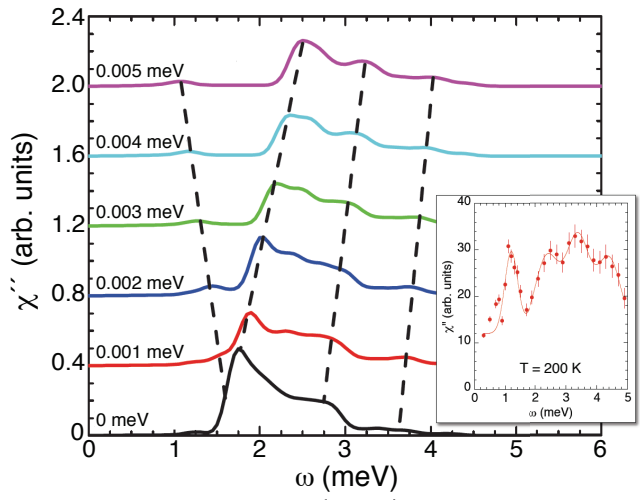}
\caption{(Color online) The averaged neutron-scattering intensity $\chi^{\prime \prime }$ at $(2\pi /a)(0.5,0.5,0.5)$ versus $\omega $ for values of $K$ from 0 to 0.005 meV.  Measurements 
at $T=200$ K are in the inset.}
\end{figure}

Earlier estimates for $K$ and $D$ were based on low-energy inelastic neutron-scattering measurements taken at the wavevector 
$(2\pi /a)(0.5,0.5,0.5)$ \cite{matsuda12}.  Because the inelastic-scattering 
cross section is quite broad, estimates based on those measurements are not as precise as the estimates described above based on spectroscopy measurements.  By using the predicted
frequencies of the $\Phi_{\pm 1}$ modes to match the primary inelastic peaks at  1.1 and 2.5 meV, Matsuda {\em et al.} estimated that $K\approx 0.005$ meV \cite{temp}.   However,
peaks in the inelastic intensity $\chi^{\prime \prime }(\omega )$ may be slightly shifted away from those frequencies due to the finite resolution of the measurements.

Averaged over a realistic resolution function and including the three sets of twins states with wavevectors along $\xp $, $\yp $, and $\zp $, $\chi^{\prime \prime }(\omega )$ is 
plotted in Fig.4 for six values of $K$ from 0 to 0.005 meV.   The inset provides the inelastic measurements at 200 K, which are almost temperature 
independent between 100 and 300 K.  For $K > 0$, the lowest two peaks are near the
$\Phi_{\pm 1}$ points.   Higher-energy peaks might be expected near the $\Psi_{\pm 1}$ and $\Phi_{\pm 2}$ zone-center modes and 
near the $\Psi_{\pm 2}$ zone-edge modes.  But due to the instrumental resolution, those peaks are smeared into broad features in $\chi^{\prime \prime }(\omega )$.

The best qualitative agreement with the inelastic measurements is obtained using $K \approx 0.004$ meV.  Below 5 meV, the 
measured $\chi^{\prime \prime }(\omega )$ contains peaks 
at 1.2, 2.4, 3.4, and 4.4 meV.  While four nearby peaks are predicted by our model, the lowest-energy peak
is too weak and the two highest-energy peaks are slightly too low compared to the experimental results.

A value for $K$ between 0.002 and 0.004 meV is consistent with the prediction $K= 0.0027$ meV obtained 
from Monte-Carlo simulations of the phase diagram in a magnetic field \cite{ohoyama11}. 
As discussed above, anisotropy produces higher harmonics $C_{2m+1 > 1}$ of the cycloid.   
Within the predicted range of $K$, $624 > (C_1/C_3 )^2 > 176$.  
Elastic neutron-scattering \cite{ramazanoglu11} and NMR measurements \cite{zalesskii02} 
indicate that $I_1/I_3 = (C_1/C_3)^2 $ is given by 500 and 25, respectively.  Because more 
precise estimates are provided by THz spectroscopy, we believe that a 
value of $K$ closer to the lower limit of 0.002 meV and a value for $(C_1/C_3)^2$ close to 500 is likely.

We conclude that the upper three peaks measured by THz spectroscopy for BiFeO$_3$ can be quantitatively predicted by a model that includes both DM and easy-axis anisotropy interactions.
Our results for the MR components in Fig.3 can be used to check the selection rules governing those three peaks.  A higher-energy EM peak at 5.39 meV is predicted with MR component $x'$.
The low-energy 2.17 meV peak measured by THz and Raman spectroscopies may be produced by additional long-range order.

Research sponsored by the U.S. Department of Energy, Office of Basic Energy Sciences, 
Materials Sciences and Engineering Division (RF) and Scientific User Facilities Division (MM), 
by the Center for Integrated Nanotechnologies, a 
U.S. Department of Energy, Office of Basic Energy Sciences user facility at Los Alamos National Laboratory, operated by 
Los Alamos National Security, LLC for the National Nuclear Security Administration of the U.S. Department of Energy (JH),
by Grants-in-Aid for Scientific Research from the Ministry of Education, Culture, and Technology, Japan (MEXT) (NF)
and by the Max Planck Society (MPG),  the Korea Ministry of Education, Science and Technology (MEST),
Gyeongsangbuk-Do and Pohang City (SM).


\begin{thebibliography}{}
\bibitem{temp} To account for the suppression of the observed moment at 200 K,
Ref.\cite{matsuda12} took $2\mu_B\sqrt{S(S+1)}= 4.11\, \mu_B$ to be about 30\% lower than its value with $S=5/2$.
Here we take $2\mu_B\sqrt{S(S+1)}=5.92\, \mu_B$
and reduce the exchange interactions accordingly.  So the estimates $J_1=6.48$ meV, $J_2=0.29$ meV, $K=0.0068$ meV, and $D=0.162$ meV from
Ref.\cite{matsuda12} are equivalent to $J_1=4.5$ meV, $J_2=0.2$ meV, $K=0.005$ meV, and $D=0.119$ meV in this paper.
\bibitem{magnetic-domain}
In a crystal with ${\bf P} \| \zp $, there are three possible orientations of ${\bf Q}$. 
The selection rules indicate that either $(0.5+\delta, 0.5, 0.5-\delta)$ 
or $(0.5, 0.5-\delta, 0.5+\delta)$ is realized in Ref.\cite{talbayev11} 
since the upper $\Psi_{\pm 1}$ mode would be inactive in field $\vh_2$
for a cycloid with $(0.5-\delta, 0.5+\delta, 0.5)$.
\end{thebibliography}
\end{document}